# Theoreticians at the Physical-Technical Institute
## (To the centennial of the Ioffe Institute)

## M. Ya. Amusia[1, 21]

[1]*A. F. Ioffe Physical-Technical Institute, St. Petersburg 194021, Russian Federation,*
[2]*The Racah Institute of Physics, the Hebrew University, Jerusalem 91904, Israel*

## Физики-теоретики ФТИ
## (К столетию Института)

**Abstract**

Dedicated to the centenary of the Ioffe Institute, the article contains the shortest review of scientific achievements of the theorists of the institute during this time. We concentrate mainly on research in the field of elementary particle physics, astrophysics, nuclear theory and atoms. To obtain very important scientific results became possible because outstanding theoreticians worked at this Institute. The high level of research persisted in spite of several mass moves of theorists - then to Kharkov, then to Moscow, then abroad. The author can testify to the atmosphere that reigned in the FTI in person, since he works at this institute since 1958. The article deals not only with research activity, but also with the famous physics schools and their outstanding cultural programs. We present also examples of extraordinary successful activity of theorists far beyond the field of their narrow specialization.

## Мирон Я. Амусья,
[1]*Физико-Технический институт им. А. Ф. Иоффе, РАН. Санкт-Петербург 194021, Российская Федерация,*
[2]*Институт Физики им. Дж. Рака, Еврейский Университет, Иерусалим, 91904, Израиль*

## Физики-теоретики ФТИ
## (К столетию Института)

**Аннотация**

Приуроченная к столетию ФТИ им. А. Ф. Иоффе, статья содержит кратчайший обзор того, что было сделано теоретиками института за это время. Речь пойдёт в основном о работах в области физики элементарных частиц, астрофизики, теории ядра и атомов. Получение важнейших научных результатов стало возможным благодаря тому, что в ФТИ работали выдающиеся теоретики. Высокий уровень исследований сохранялся несмотря на несколько массовых переездов теоретиков – то в Харьков, то в Москву, то за границу. О царившей в ФТИ атмосфере автор может свидетельствовать лично, поскольку работает в этом институте с 1958 г. В статье говорится не только об исследовательской работе, но и о знаменитых школах физики и их культурных программах. Уделяется внимание и примерам исключительно успешного выхода теоретиков за рамки своей узкой специализации.

---

[1]amusia@vms.huji.ac.il

*Сюда по новым им волнам*
*Все флаги в гости будут к нам,*
*И запируем на просторе.*
А. С. Пушкин, Медный всадник

*Мальчики, за работу. Перерыв*
*уже окончился*!
Зав отделом кадров Зинаида
Васильевна. Год, эдак, 1963-1965.
(Обращено в коридоре 2 этажа главного
здания ФТИ к бродячим теоретикам по
завершении их обеденного перерыва).

Эта заметка о теоретиках Физтеха приурочена к столетию института. Вероятно, результат был бы лучше, пиши её историк науки вообще и физики в частности. Я же не историк, и в результате желательная и беспристрастная объективность принесена в жертву собственным впечатлениям, пристрастиям, воспоминаниям, если хотите, и научным вкусам. Пишу в основном о работах в области физики элементарных частиц, астрофизики, теории ядра и атомов.

Когда-то, в доисторические времена, когда только появлялась наука физика, разделения на теоретиков и экспериментаторов не было, да и сам человек, занимающийся наукой, делал это в свободное от работы, обеспечивающей пропитание, время. Но производительность людского труда росла. Этот рост опирался на разделение труда, он же им стимулировался. Ещё Эйнштейн, а в его время наука уже была отдельным, приносящим доход и обеспечивающим приличное существование родом занятий, полагал, что теоретик – не профессия, и он должен себе зарабатывать на пропитание каким-либо конкретным делом, например, предлагал создатель теорий относительности, чинить обувь. Но нужды в этом не оказалось, и теоретики среди физиков выделились в отдельное сословие, а административно стали образовывать отделы, сектора, группы. В физтехе, у его основателя А. Ф. Иоффе теоретики с самого начала заняли достойнейшее место.

Чего от них хотели и хотят? Ведь физика – наука, в первую очередь, экспериментальная, а комбинация её с техникой ведёт, казалось бы, в сторону от абстракций и чистых размышлизмов. По счастью, так не случилось, и в Физтехе с самого начала, наряду с хаузтеоретише (во времена господства немецкого языка в физике обозначало теоретика, обслуживающего определённую экспериментальную группу или лабораторию), появились и люфт-теоретише, занимавшиеся чем им угодно. Задача теоретиков всегда сводилась к объяснению эксперимента, предложению новых экспериментов (как сказал мне один экспериментатор – простых, и способных быстро привести к славе), и обучению молодых и не очень молодых физиков этой самой физике, точнее, её передовым в настоящий момент разделам.

Никто перед теоретиками не ставит задачи – открыть новый закон физики или создать новую теорию – в доме повешенного не говорят о верёвке. Создание теории, равно как и открытие закона – нечто подразумеваемое, как правило, несбыточная мечта, тот маршальский жезл, который в абсолютном большинстве случаев совсем без толку валяется в походном ранце рядового солдата. Даже классные теоретики полагают, что работа, каждодневная и упорная, не должна определяться намерением сделать открытие. Сама каждодневная работа, изучение нового в своей области, исследование всего, что приходит в голову, разговоры с коллегами – чем больше, тем лучше, обучение других тому, что сам знаешь и себя - незнакомому – всё это должно приносить огромное удовлетворение, делающее даже каждодневную работу физика-теоретика интереснейшим занятием. А сознание того, что то, чем ты занимаешься – под силу очень немногим, позволяет относить себя к истинной элите общества – не чета всяким политикам и жуликам от бизнеса, нередко услужливыми СМИ причисляемым к элитам. И не столь важно, что в данный момент общество не относит тебя к элите. Как говаривал покойный профессор М. С. Шифрин: «От того, что тебя поместили в конюшню, не обязательно считать себя лошадью».

Вольное племя теоретиков всегда образовывало эдакого обобщённого кота, который неизменно не только «гулял сам по себе», но и указывал другим, притом не без успеха, где и как им предписывается «гулять». В целом лояльная к теоретикам администрация института периодически принимала как-то ограничивать теоретиков, подчинять их какой-то формальной «трудовой дисциплине», заставлять «приходить на работу» к определённому часу, уходить после окончания положенного «рабочего дня», брать своевременно отпуск. Тщетные усилия! Вольное племя было и остаётся вольным и не управляемым служебными административными вертикалями. В этом его сила и привлекательность.

Особенности деятельности теоретика вкупе с ощущением полной свободы, которую уже сами занятия теорией создают, определяют постоянный приток молодёжи, и ещё какой! Эти же особенности приводят к кругу общения, состоящего в основном из интеллектуально просто выдающихся людей, не только физиков, но и литераторов, музыкантов, актёров. Не удивительно, что эта специальность никогда – ни в прошлом, ни сейчас, и, я уверен, в будущем, не будет испытывать нехватки в притоке талантливой, яркой молодёжи.

На счёт института уместно относить и работы, сделанные вне его стен, уже после того, как авторы по тем или иным причинам ушли из ФТИ, где иногда и пробыли совсем немного времени. Что позволяет сделать открытие? Несомненно, в первую очередь - личный талант. Конечно же, и удача. О ней, как о главном факторе успеха, говорили, помню, на банкете, посвящённом полувекового юбилея ФТИ, знаменитые теоретики – Зельдович, Мигдал, Грибов. Однако, как писал Д. А. Гранин в романе «Иду на грозу», «удача не приходит к тому, кто ищет её вслепую».

Считаю, что даже за несколько лет работы среди таких коллег было просто невозможно не испытать их совместного влияния, которое проявлялось всю оставшуюся жизнью. Позволительно и предположение, что совсем не случайно столько ярчайших жизненных линий пересеклись, пусть и в разное время, но в одной точке пространства. Вполне правдоподобно, что это есть некая особая точка, эдакая психологическая пространственно-временная сингулярность.

Полагаю, поэтому, ошибкой то, что среди портретов Нобелевских лауреатов ФТИ в коридоре главного здания нет портрета Л. Д. Ландау, хотя, разумеется, знаю, что работы, приведшие к награждению, были написаны им, когда в ФТИ он уже не работал. Да и уходил из ФТИ он не гладко – крутой нрав молодца проявился рано и привёл к его конфликту типа «один из нас должен уйти» с директором-основателем института. Вообще, время написания работ не определяет того, когда они зародились или, точнее, пригрезились.

Я пришёл в ФТИ почти шестьдесят лет назад, в 1958. Замечу, что за эти годы нисколько не вырос вверх (слава Богу!), но никто и ничто не мешало мне продвигаться вглубь. Никаких специальных архивных данных для написания данной статьи не имел

Предыдущие, до 1958, годы истории института знаю, поэтому, лишь из мемуарной литературы и по тем устным воспоминаниям, которые дошли до меня от знакомых. Опираюсь на сведения тех, кому доверял.

К концу двадцатых, началу тридцатых годов институт имел большую группу теоретиков, состоящую из талантливейших молодых людей. Главой коллектива был крупнейший физик-теоретик СССР Я. И. Френкель (1894-1952), человек, знавший всю физику и успешно работавший во всех её областях. Экситон, позитрон, как электрон, движущийся вспять во времени, атомное ядро, как капля обычной жидкости – всех его идей не перечесть.

Для меня до сих пор остаётся загадкой, почему он, несомненно, крупнейший теоретик-ядерщик СССР, не был привлечён к атомному проекту, несмотря на все его попытки принять в нём участие. Дефектами анкеты дело не объяснишь – уж до чего дефектна была анкета Ю. Б. Харитона (бывшего Физтеховца), а был он не то что участником, но главой важнейшего направления всего проекта. Загадкой для меня остаётся и то, что избранный членом-корреспондентом АН СССР в 1929, Френкель академиком так никогда и не стал[2].

В Физтехе в конце $20^x$ – начале $30^x$ работали М. П. Бронштейн, Г. А. Гамов, Д. Д. Иваненко, В. А. Фок, зарубежным гостем был Р. Пайерлс, позднее один из руководителей английского атомного проекта сэр Рудольф (с ним, кстати, я был лично знаком). Есть замечательная фотография этих «звёздных мальчиков» из Физтеха, и среди них одна девочка – Канегиссер, из всех них избравшая Пайерлса, с которым и уехала в Англию. Кстати, недавно узнал, что брат Канегиссер в 1918 застрелил председателя петроградского ЧК С. Урицкого. А его сестру не тронули. Какие либеральные, однако, бывали времена…

Отмечу, что теоретики, в первую очередь Френкель, имели тогда прекрасные зарубежные связи, были, и становились известны за границей. Вскоре блестящие молодые, почти юные теоретики, так или иначе, но из ФТИ ушли: был расстрелян Бронштейн, на Западе остался Гамов, в Харьков перебрались Иваненко и Ландау, сменил место работы Фок.

На мой взгляд, самыми важными теоретическими работами того времени является капельная модель ядра Френкеля, на основе которой Н. Бором и Д. Уиллером был описан процесс деления атомных ядер, происходящий подобно делению капельки обыкновенной жидкости, модель Иваненко, согласно которой атомное ядро состоит из протонов и нейтронов, и количественная теория альфа-распада, т.е. вылета из ядра альфа-частиц - ядер гелия, созданная Гамовым. Замечу, что до протон-нейтронной модели ядра, физики безуспешно пытались построить ядро из протонов и электронов – единственных известных тогда элементарных частиц. Как только в 1932 Д. Чедвик открыл нейтрон, Иваненко, и месяц спустя – Гейзенберг, предположили, как оказалось, совершенно правильно, что эта частица, хоть и нестабильная – важнейший элемент ядер. Стоит помнить и о том, что цепная ядерная реакция, по аналогии с химической, была впервые рассчитана бывшими сотрудниками ФТИ, Я. Б. Зельдовичем и Ю. Б. Харитоном.

К числу выдающихся, Нобелевского уровня работ, сделанных Гамовым вне Физтеха, отношу так называемую горячую модель Вселенной (1946) и предсказание наличия генетического кода в 1954.

---

[2] У В. Я. Френкеля были некоторые документальные свидетельства, а у меня – теоретические домыслы, о роли взаимоотношений Френкеля и Ландау в описанных трудностях. Но смерть В. Я. Френкеля помешала реализации наших с ним планов совместной публикации - исследования.

Начиная со Второй мировой войны, фронт работ в области физики по всему миру начал быстро расширяться, что стимулировалось работами по созданию ядерного оружия. Это проявлялось и в увеличении числа физиков, и в строительстве многочисленных ядерных центров. В применение к СССР, уже с довоенного периода шло усиление столичной науки, рост старых и создание новых физических институтов. Ряд теоретиков из ФТИ перебрались в Москву. Я имею в виду в первую очередь таких ярких, как Я. Б. Зельдович и А. Б. Мигдал, И. Я. Померанчук, И. Е. Тамм, Г. С. Ландсберг. В Москве ещё довоенного времени начал развивать свою, основанную ещё в Харькове, и ставшую позднее всемирно знаменитой, школу теоретической физики Ландау.

В 1952 умер ещё совсем молодым Френкель. "Откачка" работников временем и Москвой создала некий вакуум среди теоретиков в институте. Однако появлялись новые люди - с полуслова всё понимающий И. М. Шмушкевич, осторожный в оценках Л. А. Слив, эрудит Л. Э. Гуревич, внешне странноватый А. И. Губанов. Они начали создавать отдел (или «загон», если угодно) теоретиков. Загоном я его называю потому, что всех теоретиков разместили в нескольких небольших комнатках полуторного этажа главного здания ФТИ. Это здание тогда ещё не перешло почти в полное распоряжение службам и службочкам, растущим, как видится мне, куда быстрее научных коллективов института.

В конце пятидесятых, как проявление общей оттепели в стране, ушли, или, точнее, ослабли анкетные препоны в приёме научных сотрудников, что в первую очередь сказалось на пополнении теоретиков. В ФТИ пришли А. З. Долгинов, В. Е. Голант, В. Н. Грибов, В. М. Шехтер, В. И. Перель, Г. М. Элиашберг, Д. А. Варшалович, А. А. Ансельм, С. В. Малеев, Р. Ф. Казаринов, В. Г. Горшков и ряд других. Главным поставщиком молодёжи служил, естественно, физфак ЛГУ, где совсем не формально отдел теоретической физики возглавлял В. А. Фок, в некотором смысле сам «уроженец» Физтеха. И хотя общая формула Слива, согласно которой «в науку пошёл середняк» правильна, молодыми теоретиками ФТИ были отнюдь не только середняки.

Вновь принятых старших лаборантов и аспирантов администрация не оставляла без опеки и внимания, а перво-наперво отправляла на проверку реальным делом. Это видно на Фото. 1.

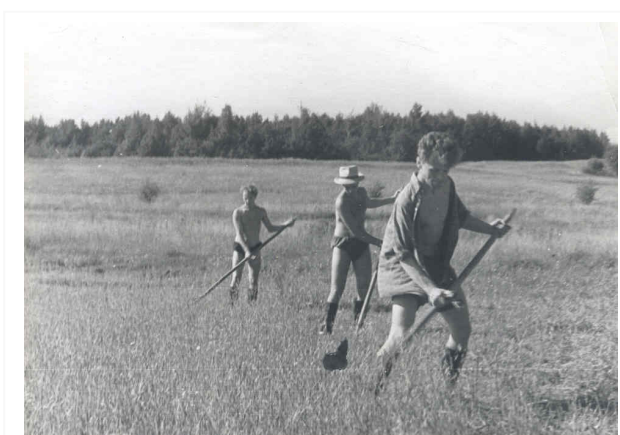

Фото 1. 1958 г. Первый «пропуск» в ФТИ, слева направо: С. Шерман, автор, ?.

Начали восстанавливаться порванные войной международные научные связи. Здесь первым стал Слив, у которого возникло тогда крайне редкое заочное сотрудничество с Институтом Нильса Бора в Копенгагене, конкретно с О. Бором. Слив туда поехал, и на месте, выслушав четыре лекции самого Бора по физике, понял, что тот никакой не законченный идеалист, а просто натуральный диалектический материалист, не чета ещё развившейся по привычке философской околонаучной шпане, вроде Львова. О своих находках Слив доложил на общеинститутском семинаре, полностью ошеломлённом его сообщением. Кстати, на меня О. Бор, с которым я был лично знаком, произвёл особое впечатление ещё и тем, что некоторые письма писал Сливу по-русски. Я почитал и решил – если это русский, то я могу писать по-английски. И начал этим заниматься.

Материальным результатом поездки Слива в Копенгаген стал ответный визит Б. Моттельсона, будущего Нобелевского лауреата. Он ознакомился с молодыми теоретиками, дав возможность каждому из них выступить перед ним с пятиминутным сообщением. По результатам этой беседы я получил за подписью Н. Бора письмо с приглашением провести год в Копенгагене. Не пустили, а письмо куда-то дели. Примерно через десять лет, уже от О. Бора, сходное приглашение получил В. Н. Ефимов, но его «силовики» буквально вывели из самолёта. Я это пишу к тому, что злопамятность отнюдь не вреднее памяти короткой.

Вообще, иностранные теоретики были нередкими гостями в ФТИ. Упомяну, естественно, лишь некоторых: Нобелевских лауреатов П. А. М. Дирака, С. Томонагу, Ю. Швингера, Д. Бардина, О. Бора, а также известнейших профессоров У. Фано, Р. Пайерльса (сэра Рудольфа Эрнста), Ф. Бёрка, Дж. Брауна, В. Грайнера.

Вспоминаю, как прямо на институтском семинаре Грибов с Дираком поспорили о знаке перед массой в уравнениях Дирака, и Грибов выиграл спор! Вспоминаю, как оценивая уравнения Элиашберга в теории сверхпроводимости, Д. Бардин поздравил Гуревича, отметив, что это честь – иметь такого сотрудника. Как и до войны, теоретики ФТИ занимались практически всеми областями физики – от твёрдых тел, жидкости, газов, плазмы, до теории элементарных частиц и фундаментальных проблем строения Вселенной. Это удобно и полезно - иметь под одной крышей всех специалистов сразу. Можно, если возникает вопрос или идея, не рыться в литературе, не заглядывать даже во всезнающую Википедию, а просто позвонить приятелю, и получить консультацию на самом что ни на есть высоком уровне.

В. Н. Грибов (Фото 2) был, несомненно, самым ярким и влиятельным из молодых. Всемирную известность он приобрёл за работы, которые в целом можно называть реджистикой – по имени итальянского теоретика Т. Редже, первым исследовавшим амплитуду рассеяния как функцию углового момента, рассматриваемого континуальной комплексной переменной. Редже, с которым мы познакомились в Новосибирске, изучал лишь нерелятивистский случай. Грибов привнёс этот подход в физику элементарных частиц, смело предположив, что асимптотика амплитуды по энергии столкновения определяется первым полюсом Редже. Тогда казалось, что асимптотика достигнута, или вот-вот будет достигнута, на эксперименте. Опыт показал, что это не так. Но реджистика уже стала важнейшим направлением в физике элементарных частиц.

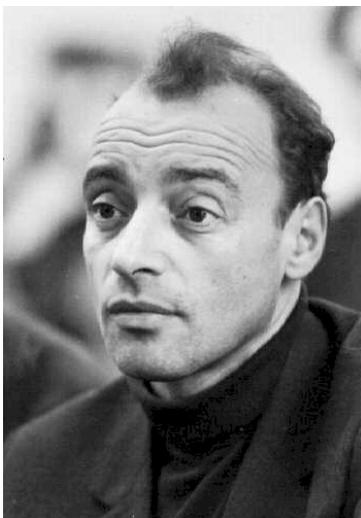

Вообще, по глубине знаний, кругозору, да и важности собственных работ, даже по внешности – в первую очередь блеску глаз, мне, да и не только мне, Грибов казался новой реинкарнацией Ландау. Помню, как поразила меня символичность сочетания номера

Фото 2. В. Н. Грибов

страницы -1972, и года публикации – 1971 – одной из его работ - статья буквально опережала своё время!

Формула Грибова, связывающая сечение рассеяния пи-мезона на протоне $\sigma_{\pi p}$ с сечением рассеяния пи-мезона на пи-мезоне $\sigma_{\pi\pi}$ и протона на протоне $\sigma_{pp}$ в пределе высоких энергий столкновений, представляется очень простым и красивым выражением:

$$\sigma_{\pi p}^2 = \sigma_{\pi\pi}\sigma_{pp}.$$

Она хорошо описывает экспериментальные данные даже в той области, где основания, приведшие к её выводу, уже представляются несправедливыми.

Такие формулы, если правильные, вполне могут высекаться на надгробных камнях их авторов – по примеру выражения для энтропии на памятнике Л. Больцману.

Грибов баллотировался сразу в академики, минуя член-корство, но ему не хватило буквально одного голоса. Учитывая прямоту и открытость его научных суждений, некое противостояние с Н. Н. Боголюбовым и его школой, следовательно, наличие врагов, происшедшее не удивляло.

Грибов был абсолютным авторитетом в вопросах физики в ФТИ и далеко за его пределами, в том числе и вне рамок тех областей, где он непосредственно работал. Время, однако, безжалостный и объективный судья. Оно показало, что не все оценки Грибова, как и Ландау, пусть стремительные и чёткие, были правильными. Так, как оказалось, он зря отвергал работу Глинера (см. ниже), и, как показало время, напрасно считал неравенства Белла, сейчас знаменитые, бессодержательными, а проверку квантовой механики с их помощью, ненужной.

В книге «Неизбежность странного мира», вышедшей в 1961 и очень популярной в своё время, писатель Д. Данин, сам по образованию физик, упоминает об особой «улыбке Грибова», и говорит о нём и его ближайших сотрудниках как о тех, кто создаёт картину этого мира.

Когда только появилось представление о кварках как об основных элементах, из

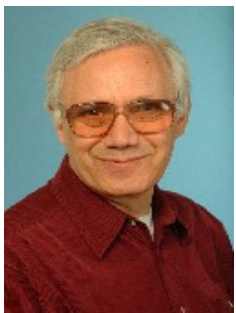

которых построены элементарные частицы, специалисты сначала с опаской приспосабливались к этим новым объектам физики элементарных частиц. Однако два физтеховца - Е. Левин (Фото 3) и Л. Франкфурт (Фото 4), подошли к делу просто. Они буквально за пани брата обращались с новыми объектами. Действительно, рассудили они, раз мезон состоит из кварка и антикварка, а протон или нейтрон – из трёх кварков, сечения

Фото 3 Е. М. Левин

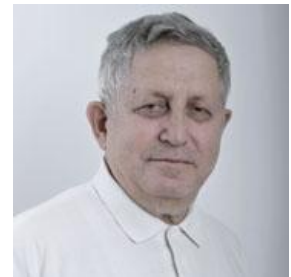

Фото 4.
Л. Л. Франкфурт

их взаимодействия с самими собой и другими объектами должны относиться, соответственно, как числа входящих в них компонент, т.е. как 2:3. Соответствующая работа была опубликована в Письмах ЖЭТФ в 1965.

Это, ставшее знаменитым «2/3», и более сложные соотношения, полученные на основе столь простой картины, на удивление хорошо описывали реальность. Хотя она, эта реальность, и оказалась куда сложнее картинки, согласно которой «элементарная частица» подобна тоненькой и ни на что не влияющей банке, в которой свободно болтаются кварки и антикварки. Предложенная метода завоевала широкое признание и получила название «кваркового счёта».

В 1965 в ЖЭТФ вышла статья Э. Б. Глинера «Алгебраические свойства тензора энергии-импульса и вакуумоподобное состояние вещества», ставшего сотрудником ФТИ только в 1964. В ней впервые он дал физическую интерпретацию космологической постоянной Эйнштейна и выдвинул гипотезу о физической природе Большого Взрыва. По Глинеру (Фото 5), вначале во Вселенной был вакуум, описываемый космологической постоянной. Из первичного вакуума рождалось

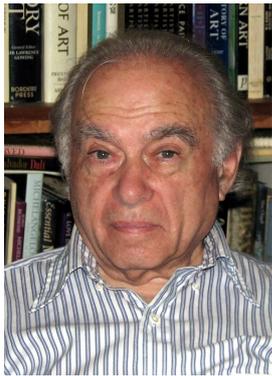

Фото 5. Э. Б. Глинер

вещество, и оно расширялось под действием антигравитации вакуума. Так возникло наблюдаемое космологическое расширение.

Эта работа стала важнейшей для понимания того, что происходило во Вселенной в первые, немыслимо короткие мгновения её существования. Она ввела в обиход космологии понятие отрицательного давления, и в то же время позволяла понять, почему такое давление не приводит к коллапсу системы «Вселенная» под действием обычных гравитационных сил.

Работу поддержал А. Д. Сахаров, но тогда его участие было скорее отрицательным фактором, затруднившим защиту кандидатской диссертации, которую удалось провести только в Тарту, а не в ФТИ. Известности работ Глинера способствовало энергичное развитие этой ветви космологии, которая связана с именами Гута и А. Линде. Со временем, Глинер был признан основоположником этого важнейшего направления. Однако это «со временем» надо было ещё прожить. Глинер пришёл в ФТИ после окончания ЛГУ, которому предшествовало и участие в Великой Отечественной войне, на которой заслужил ордена, был трижды ранен, потерял руку, и тюрьма. Военные заслуги не помешали осудить его за «неправильные» разговоры, и он долгое время провёл в заключении.

Совсем уже немолодой, он оставался младшим научным сотрудником, что, помимо скромности зарплаты, ещё и задевало самолюбие. Здесь помощь могли оказать коллеги, но они, за редким исключением, либо не могли помочь, либо стояли в стороне. Прав был В. Л. Гинзбург, написавший в УФН в 2002: «Я считаю, что мы в большом долгу перед Э. Б. Глинером». О таких долгах стоит помнить, равно как и знать о них, и молодым, чтобы не ошибаться в жизни. И каждый раз, когда мы встречаемся с «раздувающейся Вселенной», «инфляционной гипотезой», уместно помнить, что исходные идеи в этом направлении, равно как и важное участие в её развитии принадлежат Глинеру, который сейчас живёт в США.

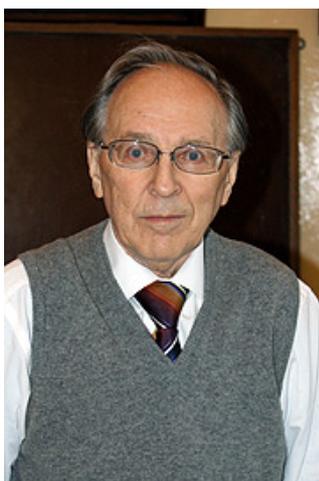

Фото 6. Д. А. Варшалович

Очень яркая фигура среди теоретиков Д. А. Варшалович (Фото 6), которого судьба и собственный интерес двигали из ядерной физики в астрофизику так, что на каждом шаге этого пути оставались впечатляющие отметины. Так, им предсказано позднее обнаруженный эффект выстраивания спинов атомов, ионов и молекул в космической среде под влиянием резонансных анизотропных потоков излучения. Большой научной смелости требовала работа, показавшая, вопреки абсолютно устоявшимся представлениям, инверсную заселённость уровней возбуждения молекул в космической среде, которая оказалась своего рода гигантским мазером. Варшалович с сотрудниками предсказал существование в космосе молекулы HD, позднее обнаруженной в далёких космических облаках. Излучение этой молекулы имеет красное смещение 2-3, т.е. идёт от источников, отстоящих от нас на 12 – 13 миллиардов световых лет!

Варшалович и его сотрудники, из которых особо упомяну молодого члена-корреспондента РАН А. В. Иванчика, нашли, по данным спектров удалённых космических объектов – квазаров, верхние на сегодняшний момент границы изменения во времени и пространстве таких констант, как постоянная тонкой структуры $\alpha$, т.е. безразмерная комбинация заряда электрона $e$, скорости света $c$ и постоянной Планка $\hbar$ $\alpha \approx e^2 / \hbar c \approx 1/137$, и отношение массы протона к массе электрона. Эти изменения составляют не более, чем примерно $10^{-5}$-$10^{-4}$ за всё время, прошедшее после Большого взрыва, т.е. за примерно 14 миллиардов лет. Хотя и крайне малые, эти изменения, если их удастся обнаружить, станут важнейшими для построения всей физической картины мира.

В 1975 была опубликована книга Варшаловича Д. А., Москалёва А. Н. и Херсонсккого В. К. Квантовая теория углового момента. Вскоре опубликованная на английском, она стала, и в большой мере остаётся, уникальным пособием (буквально настольной, что неоднократно видел сам) для множества теоретиков во всём мире. Она уникальна тем, что, авторы привели в единую систему огромное количество уже имевшихся в литературе формул, существенно различавшихся используемыми обозначениями. В результате, в литературе сосуществовали сходные выражения, различающиеся друг от друга по величине и фазе. Это затрудняло их использование, поскольку часто приходилось брать части используемых соотношений из разных статей или книг. Все, даже самые сложные известные формулы авторы проверили и перевывели. В них было обнаружено немало ошибок. Книга включила также ряд новых, не известных ранее соотношений. Если правильно понимаю смысл старинного слова «трактат», то оно в полной мере может быть применено к труду Варшаловича с соавторами.

У Варшаловича есть, как и была у Грибова, одна поразительная черта: о чём бы ни завести с ним разговор, он всегда к нему готов. Обсуждаемый вопрос оказывается тщательно продуман, а ответы заставляют во многом по-новому взглянуть на то, что, собственно, обсуждается. Вывод из этого можно сделать один – сказанное Варшаловичем есть предмет долгих и глубоких размышлений обо всём, что касается физики – всей физики, в целом, а не только тех её разделов, о которых он публиковал или публикует свои статьи.

Говоря об астрофизике, на память приходит и созданное природой гигантское, размером в километры, ядро, наподобие атомного, но состоящее в основном из нейтронов. Я имею в виду нейтронные звёзды. Физика ядра и конденсированного вещества подсказали вопрос – а могут ли нейтронные звёзды быть сверхтекучими? Помню оживлённые дебаты на эту тему в шестидесятых, помню тогдашние выводы – сверхтекучесть, если вообще возможна, то только в тонком поверхностном слое. Время споры разрешило. Было установлено, что нейтронные звёзды сверхтекучи в объёме, и сохраняют это свойство при весьма экзотических даже для этих экзотичнейших из объектов, условиях. В появлении этого ответа огромную роль сыграл физеховец Д. Г. Яковлев с сотрудниками.

Важным направлением в работе теоретиков ФТИ уже давно было изучение проблемы трёх тел, которая в обычной и квантовой механике следует по сложности сразу за проблемой двух тел.

Пригодное для аналитических и численных расчётов уравнение теории трёх тел было написано в ФТИ Г. В. Скорняковым и тогдашним его научным руководителем физеховцем К. А. Тер-Мартиросяном сразу для межчастичных сил нулевого радиуса[3]. Оказалось, что для любых, даже слабых, сил притяжения, оно приводит к странному, казавшемуся нефизическим, результату – бесконечному числу дискретных связанных уровней. Происхождение этой бесконечности было совершенно не ясно.

---

[3] Для любых сил соответствующие уравнения были получены Л. Д. Фадеевым и вошли в науку под его именем.

В 1970 <u>В. Н. Ефимов</u> (Фото 7), продолжая и развивая нашу с ним работу по энергии газа твёрдых шаров, обнаружил новый эффект, вскоре получивший его имя. Оказалось, что в системе трёх частиц, взаимодействующих силами малого радиуса $r_0$, с так называемой длиной рассеяния $a$, фактически, эффективным размером частицы с точки зрения её столкновения с другой частицей, возникают специфические уровни энергии, от радиуса действия сил и их радиальной формы не зависящие. Их назвали «уровнями Ефимова». Число этих уровней $N$ определяется замечательной по простоте и изяществу формулой:

$$N \approx \frac{1}{\pi} \ln a / r_0.$$

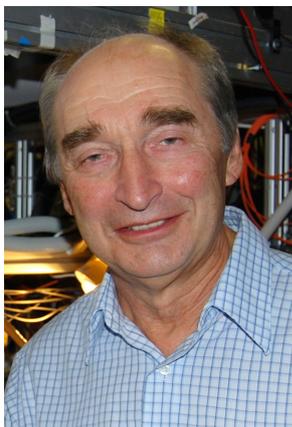

Фото 7. В. Н. Ефимов

Красива и формула для энергии семейства уровней энергии Ефимова, близкая к следующему весьма простому выражению:

$$E_n / E_{n+1} \approx \exp 2\pi.$$

Видно, что при нулевом радиусе действия сил $r_0 \to 0$, $N \to \infty$, в согласии с результатом, получаемым с помощью уравнений Скорнякова – Тер-Мартиросяна. Все уровни представлялись наблюдаемыми, и имели физический смысл. Но в природе не нашлось примеров достаточно малых значений $r_0$ и больших $a$. Спустя более, чем сорок лет после предсказания за дело взялась квантовая оптика. Регулируя частоту лазера, удалось менять длину рассеяния атомов, конкретно - цезия, достигая при этом очень больших значений, положительных и отрицательных, длины рассеяния. Открылся целый «мир Ефимова», управляемый необычными законами. Словом, думаю, что самому Ефимову вскоре может понадобиться чёрный фрак или смокинг.

Это не только моё мнение – иначе не вошёл бы он, один из немногих, в интересный и очень короткий список «<u>"50 People Who Deserve a Nobel Prize". The Best Schools</u>». В разделе «физика» этого списка включено десять имён, образующих достойную компанию. Это Якир Ааронов (эффект Бома-Аронова), Алейн Аспект, («запутывание» фотонов), Виталий Ефимов, («состояния Ефимова», иллюстрирующееся старинным символом - тройкой переплетённых колец, из которых убери любое - оставшаяся пара распадается – Фото. 3), Алан Гут («космическая инфляция» или «раздувающаяся Вселенная»), Л. Вестергад Хау (полная остановка, вплоть до нулевой скорости, света), Петер Ниггс (бозон Хиггса), Лев Питаевский (уравнения Гросса-Питаевского). Один из списка, Хиггс, в 2014 уже стал лауреатом. Как говорят, «первый пошёл».

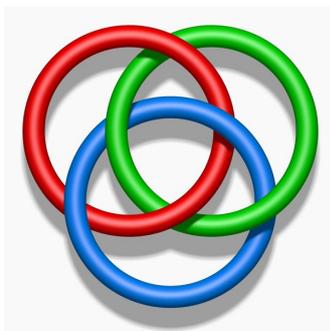

Кто заинтересуется – советую прочитать очень эмоциональное сообщение о полном экспериментальном подтверждении в 2014 факта существования нового состояния вещества, наблюдающегося, при крайне низких температурах. Символически, «состояния Ефимова» представлены на Фото 8 – расцепи одно кольцо, и вся тройка развалится. Радиус обнаруженного состояния тройки составляет тысячу атомных радиусов! На таких расстояниях

Фото 8. Символически – «состояние Ефимова»

удерживающие их вместе силы крайне слабы, а энергия связи мала, что затрудняет опыт. Соответствующая статья называется: «[Физики доказали удивительное правило троек](#)», где говорится даже о мистике числа три (при том, знакомые по СССР «тройки» не упоминаются): над открытием работали три группы, в трёх странах, изучая сплетение трёх тел! Подводя некий итог этой теме, в 2017 появилась статья с примечательным названием «[Физика Ефимова: обзор](#)». Ефимов работает в Университете Сиэтла. Он – блестящий лектор, что видно из его выступлений по физике, выложенных в YouTube. А пока что международная группа экспертов выбрала первых лауреатов премии [Фаддеева](#) 2018 года: В. Ефимова, «*За теоретическое открытие ряда слабосвязанных трёхчастичных квантовых состояний, известных как состояния Ефимова*» и Р. Гримма, из Австрии, «*В знак признания его новаторских экспериментов, подтверждающих эффект Ефимова*». [Награждение](#) состоялось на конференции в Кане 11 июля 2018.

В конце $50^x$ - начале $60^x$ казалось, что всё об аналитических свойствах амплитуд рассеяния разных частиц друг на друге в функции энергии столкновения, рассматриваемой как комплексная переменная, хорошо известно. Именно, они содержат в комплексной плоскости лишь простые полюса и разрезы. Знание аналитических свойств амплитуды рассеяния позволяет строить так называемые дисперсионные соотношения, связывающие её реальную и мнимую части. А вместе с ними и связывать различные измеряемые на опыте характеристики процессов рассеяния атомных ядер, да и более сложных объектов, из них состоящих, непосредственно, минуя разные модели, описывающие межчастичное взаимодействие.

Казалось, что в этой области, если в качестве сталкивающихся частиц взять электрон и атом, уж точно ничего интересного не найдёшь. Данному вопросу посвящён параграф в «Квантовой механике» Ландау и Лифшица, где, со ссылкой на Л. Д. Фаддеева, приведено дисперсионное соотношение амплитуды рассеяния электрона на атоме. Согласно Книге, в амплитуде рассеяния есть лишь полюс первого порядка и разрез.

Однако учёт обмена налетающего электрона с атомным электроном кардинально меняет дело. Важную роль играет и то, что силы между электронами дальнодействующие, кулоновские, а не короткодействующие, ядерные. Учёт обеих факторов полностью меняет аналитические свойства амплитуды. Вместе с М. Ю. Кучиевым (Фото 9), пришедшим в ФТИ в 1974 ко мне в аспирантуру, мы показали в 1979, что амплитуда рассеяния электрона на атоме водорода $f_{eH}$ как функция энергии столкновения $E$ имеет полюс третьего, а не первого, порядка $f_{eH}(E \to E_H) \sim 1/(E - E_H)^3$, где $E_H$ есть энергия связи атома водорода.

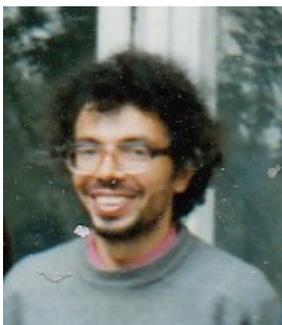

Фото 9. М. Ю. Кучиев

Ещё более неожиданной оказалась ситуация для произвольного атома. Как показал Кучиев, в этом случае справедливо следующее выражение для амплитуды электрон-атомного рассеяния

$$f_{eA}(E \to E_A) \sim (E - E_A)^{2+|E_A|/|E_H|},$$

где $E_A$ есть энергия связи электрона в атоме-мишени. Следовательно, сингулярность здесь вообще иррациональна. Этот результат может оказаться важным и в квантовой хромодинамике, и в описании рассеяния электронов на поверхностях твёрдых тел.

Успех сопутствовал Мише и в решении другой проблемы. Уже первые исследования взаимодействия лазерного излучения высокой интенсивности с атомами показали, что здесь играет роль какой-то неизвестный механизм. Этот механизм увеличивает вероятность многоэлектронной фотоионизации облучаемого атома на несколько десятков порядков, по сравнению с ожидаемой вероятностью, получаемой исходя из обычной теории. Почти сразу стало ясно, что достаточно интенсивному лазерному излучению удаётся удалить из атома электрон и в том случае, когда энергия одного лазерного фотона много меньше энергии, необходимой для ионизации атома. Это означало, что происходит многофотонная ионизация.

Вскоре интенсивность лазерного излучения и соответствующая ей напряжённость электрического поля превзошла напряжённость атомного поля в водороде. Это был важнейший рубеж в истории человечества, как в своё время первый ядерный взрыв, оказавшийся «ярче тысячи солнц». Сейчас напряжённости в лазерном поле уже на много порядков превосходят атомные. Обнаружены процессы, требующие поглощения многих сотен или даже тысяч лазерных фотонов.

Механизм взаимодействия низкочастотного излучения с атомами и любыми атомоподобными объектами – кластерами, фуллеренами, эндоэдралами, был открыт Кучиевым в его статье «Атомная антенна» в 1987. Он обратил внимание на то, что ионизованный электрон, колеблясь в лазерном поле, приобретает энергию $E$, определяемую соотношением:

$$E \sim I/\omega^2 ,$$

где $I$ есть интенсивность лазерного поля, а $\omega$ – его частота. Уже для $I$ порядка поля в атоме водорода, т.е. $10^{16}$ Ватт/см$^2$ и энергии лазерного фотона в 0.1эв, $E$ достигает двух миллионов эв. С этой энергией электрон, колеблясь относительно атома, из которого выбит, возвращается к нему. При этом у него достаточно энергии, чтобы выбить уже несколько электронов из атома, или стать источником фотонов с энергией, многократно превышающей энергию лазерного фотона $\hbar\omega$. Сейчас данный механизм, часто называемый «обратным рассеянием», общепринят. Он уточняется, и в рамках основной идеи появляется целый ряд ответвлений. В данной области работает много исследователей. В целом, «обратное рассеяние» служит для описания огромного экспериментального материала. Ставятся специальные опыты для его обнаружения, и оно используется для объяснения множества проведенных опытов. Но, увы, как правило, без ссылок на Кучиева. Сейчас он сам продолжает успешно работать в Сиднее, но по другой тематике.

Знаменитые уравнения Хартри-Фока (ХФ) выводятся, исходя из требования - обеспечить минимум энергии атома на базе его простейших волновых функций, представляемых антисимметризованным произведением одноэлектронных волновых функций. Многие десятилетия они использовались как основные в расчётах структуры атомов и процессов, происходящих с их участием. Это были первые уравнения, описывающие квантовую систему многих тел. Их область применений выходит сейчас далеко за рамки физики атома, включая ядерную физику, описание твёрдых тел и конденсированных веществ.

Уравнения ХФ удобны в записи, но не просты в решении. Кроме того, в течение довольно длительного времени практически не обсуждался вопрос о величине и значимости поправок к тому, что учитывается в рамках этих уравнений, и именуется словами «электронные корреляции».

Идеальным аппаратом для выхода за рамки ХФ служит общая теория многих тел и диаграммная техника. Она была заимствована из квантовой электродинамики, и привнесена в теорию многих тел работами в первую очередь Мигдала и В. М. Галицкого, посвящённых пространственно-неограниченным и однородным по плотности частиц объектам.

Но в атомах, равно как и в простых молекулах, важная роль неоднородности системы просто несомненна, что сразу поставило вопрос о проведении трудоёмких вычислений на компьютерах. В 1964, в ФТИ, я (Фото 10) начал работу по выяснению роли электронных корреляций в атомных процессах, в первую очередь, фотоионизации атомов и неупругом рассеянии быстрых электронов на них. Первоначально не задумывался над вопросом, сколько на это уйдёт времени в человеко-годах. Оказалось, однако, что со временем тематика только расширяется, затрагивая всё новые и новые процессы. Важнейшим уточнением ХФ стало предложенное и разработанное нами приближение случайных фаз с обменом (ПСФО, или, в английской версии – RPAE).

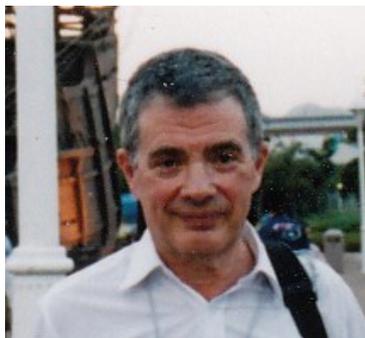

Фото 10. М. Я. Амусья

Постепенно была создана неформальная группа физиков, с переменным числом участников, иногда доходящим до 25 человек. Их потребности в вычислениях обеспечивали программы, создаваемые одним программистом – Л. В. Чернышёвой (Фото 11), и образующие единый комплекс, допускающий непрерывное расширение по мере рассмотрения новых физических задач. Современная версия этого уникального комплекса, обнимающая не только атомные, но и молекулярные программы, а также программы, необходимые для расчётов процессов, с участием эндоэдралов, собрана в книге М. Я. Амусья, С. К. Семенова и Л. В. Чернышевой, *АТОМ-М. Алгоритмы и программы для исследования атомных и молекулярных процессов*, Издательство «Наука», Санкт-Петербург, 2016, 551 стр.

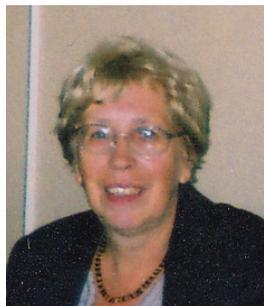

Фото 11.
Л. В. Чернышева

В рамках этих исследований в 1967 было показано, что так называемый гигантский резонанс в сечении фотопоглощения ряда атомов есть коллективное возбуждение, в котором участвуют, по меньшей мере, все электроны ионизуемой многоэлектронной подоболочки. В 1972 были предсказаны, и позднее обнаружены на опыте, интерференционные минимумы, возникающие благодаря взаимодействию электронов, принадлежащих разным оболочкам, что приводит к межоболочечным резонансам. Исследование свойств и различных проявлений этих резонансов стало предметом многочисленных последующих исследований в целом ряде атомных лабораторий мира. Было предсказано также существование поляризационного или «атомного» механизма генерации электромагнитного излучения и многое другое. Основные наши результаты, полученные в многочастичной теории атомных процессов, приведены в книге M. Ya. Amusia, L. V. Chernysheva and V. G. Yarzhemsky, *Handbook of theoretical Atomic Physics, Data for photon absorption, electron scattering, and vacancies decay*, Springer, Berlin, 2012. pp. 812.

Остановлюсь несколько подробнее на поляризационном механизме генерации электромагнитного излучения, описанном нами впервые в 1976. Обычное тормозное излучение (ОТИ), возникает при рассеянии электрического заряда статическим полем атома, ядра, молекулы и т.п. Излучает в таком рассеянии заряд. В «поляризационном» тормозном излучении (ПТИ) источником излучения может быть и нейтральная частица, внутреннее распределение зарядов в которой меняется в процессе столкновения. В результате в ней наводится переменный во времени (как правило, дипольный) электрический момент. Именно этот момент и становится источником электромагнитного излучения. Данный механизм приводит к излучению и при столкновении нейтральных частиц, притом, не только атомов. Так, ПТИ должно возникать и в столкновениях, например, нейтрино и нейтрона. Генерируется ПТИ и при столкновениях макрообъектов. Исследование ПТИ продолжается в ряде экспериментальных и теоретических групп в разных странах и сейчас.

Выше я говорил о предсказании Варшаловичем выстраивания спинов атомов и ионов в космической среде под влиянием направленного излучения. Оказалось, что спины фотоэлектронов, вылетающих из атома под действием падающего на него электромагнитного излучения, также выстроены. Этот эффект был предсказан, и следствия его обсуждены в работах Н. А. Черепкова (Фото 12) в <u>1972</u> и <u>1973</u> гг. Черепков показал, что электроны, вылетающие из определённой подоболочки атома, могут быть под определёнными углами испускания полностью поляризованы, даже если ни поток ионизующих фотонов, ни атом-мишень не поляризованы, т. е. избранным направлением не обладает. Предсказание было подтверждено

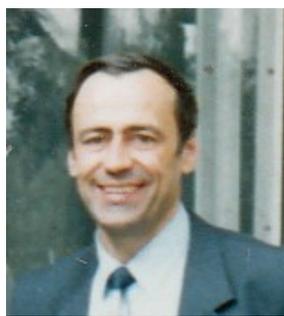
Фото 12. Н. А. Черепков

на опыте. Изучение поляризации электронов, удаляемых с помощью фотонов, стало одним из важнейших методов в исследовании структуры атомов и молекул.

Исследования многоэлектронных эффектов продолжаются и расширяются, включая новые эффекты в атомах и новые объекты исследования, куда теперь входят кластеры металлов, фуллерены и эндоэдралы. Новые объекты во многих отношениях подобны атомам, только гигантским, имеющим сотни, а иногда и тысячи электронов. В то же время, эти объекты обладают многими свойствами, например, гигантским резонансом в вероятности поглощения фотонов, характерным не только для атомов, но и атомных ядер. Процесс фотоионизации исследуется теперь и с помощью лазеров, длительность светового импульса которого составляет всего $10^{-17}$-$10^{-18}$ секунды, что позволяет получить представления о протекании атомных процессов во времени. Примечательно, что описание того, как процесс ионизации атомов разворачивается во времени, достигают, используя наш комплекс вычислительных программ АТОМ-М.

Я говорил уже о том, что теоретику присуще желание самому учиться, и учить других. Это желание привело к организации в 1966 Зимней школы физики ФТИ, которая первоначально включала, помимо ядра и элементарных частиц, ещё твёрдое тело и конденсированное состояние. Как Зимнюю школу по физике ядра и элементарных частиц, её «унаследовал» ЛИЯФ, ставший затем ПИЯФ, а позднее обратившийся в часть Национального исследовательского центра «Курчатовский институт». «Шапки» менялись, а школа по целям, строю и духу оставалась той же. Существующая более полувека, она для своего описания заслуженно требует целой книги, которая, уверен, будет написана.

По широте и глубине охвата материала, по уровню лекторов – лучших научных работников СССР и, позднее, РФ, по остроте дискуссий, которые шли нередко далеко за полночь, она, пожалуй, не имеет себе равных в мире. Давно обрела она буквально международное признание. Особо стоит отметить труды школы, собравшие все лекции и представившие их на суд научной общественности – сначала в СССР, а затем и за границей. Первоначально созданная для повышения научной квалификации экспериментаторов, она вскоре стала школой для всех научных работников.

Отличительная черта этой школы – использование заметной части свободного времени для повышения образования участников школы в области литературы и искусства. Первые десятилетия её совпали со временем, о котором говорили «что-то физики в почёте, что-то лирики в загоне». И физики помогали им вылезать из «загона», в котором, как я упоминал в самом начале этой заметки, сидели сами, что ценилось «лириками» высоко. А среди гостей – лириков бывали Г. Товстоногов, Б. Окуджава, А. Володин, С. Юрский, К. Лавров, Н. Симонов – всех звёзд не перечислить. Недавно напомнил А. Сокурову, как он, ещё почти никому не известный, показывал на школе свои фильмы, «Одинокий голос человека» и «Сонату для Гитлера», и при этом явно нервничал по пустякам.

Как-то Товстоногов, расчувствовавшись, пригласил всех к себе в театр. Приглашение приняли, и обе стороны долгие годы выполняли взятые на себя обязательства – физики в пристойном числе ходили в театр, а Товстоногов обеспечивал их билетами. Стены залов и коридоров школы использовались для экспозиций работ художников, которые не имели никакого официального признания. Зато физики признавали их уже фактом приглашения к себе. Сейчас некоторые из гостей школы представлены в музеях и галереях мира. Это общение придавало Зимним школам дополнительную притягательность, которую не забыть. Теперь уже «лирики» в помощи физиков не нуждаются, да те вряд ли могут её оказать.

Теоретиков отличает не только желание учить и учиться, но и стремление применять законы физики, известные им, а иногда и открываемые с их помощью, к изучению проблем в областях, на первый взгляд, прямо к физике не относящихся. Например, в медицине, экономике, политике. Приведу пару особо впечатливших меня примеров: работы В. Г. Горшкова (Фото 13) по экологии и М. И. Дьяконова (Фото 14) в области «квантового компьютинга», точнее, его отрицания.

Горшков получил очень интересные результаты, иллюстрирующие важность сохранения окружающей среды, применяя к её описанию довольно общие соотношения, полученные в физике. Он пришёл к важному выводу об определяющем влиянии естественных экосистем (биоты) на поддержание параметров окружающей среды в пределах, допускающих само существование жизни на Земле.

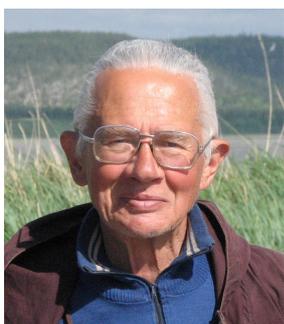

Я - свидетель зарождения этого направления, неоднократно слушал доклады Горшкова, и, не являясь его адептом, могу оценить, какую убедительность придают словам количественные оценки и расчёты, основанные на установленных законах физики. В процессе написания данной

Фото 13. В. Г. Горшков

статьи, прочёл с большим удовольствием его интервью от 2008 г, где развиваются физико-химические представления о движущих силах континентального влагооборота.

Дьяконов, видный специалист по теории твёрдого тела, сейчас профессор университета в Монпелье, уже более пятнадцати лет выступает со всё более резкой критикой так называемого «квантового компьютинга», став одним из наиболее известных и убедительных «скептиков» в данной области. Это выступление против лидирующего направления, называемого ещё мейнстримом - воплощение достойнейшего из принципов - «Иду на грозу».

Идея «квантового компьютинга», сейчас важнейшего и отлично финансируемого направления современной теории компьютеров, восходит к работам Ю. И. Манина 1980 и Р. Фейнмана 1981 гг. Казалось, что на основе развития этих идей - записи информации в виде суперпозиции состояний квантовой системы с её огромным многообразием возможностей, удастся несказанно повысить быстродействие и увеличить объём памяти компьютеров. Это направление привлекает очень большое число исследователей, и под него создаются новые лаборатории и институты.

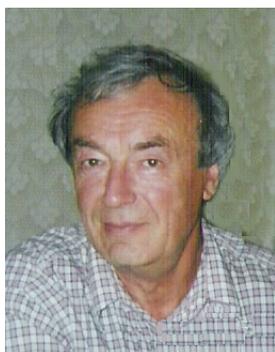

Дьяконов же доказывает, что из-за неизбежных шумов в физических устройствах обещанная быстрота вычислений не достижима. Все знаменитые теоремы в этой области, отмечает он, основаны на аксиомах, в принципе нереализуемых в мыслимых физических устройствах. Это утверждение сродни замечанию Ландау, что все строгие теоремы в квантовой механике либо неверны, либо бессодержательны.

Дьяконов отмечает также, что, произойди чудо из чудес, и квантовый компьютер был бы создан, в принципе для него не нашлось бы подходящих задач. Он анализирует причины невыполнения взятых на себя «компьютерным сообществом» временных обязательств, и отмечает, что в рамках основополагающего в этой области алгоритма Шора, который должен был совершать чудеса в краеугольной вычислительной

Фото 14.
М. И. Дьяконов

задаче – факторизации очень больших чисел, пока удалось лишь показать, что 15=5x3 и 21=7x3! Замечу, что те несколько статей Дьяконова на эту тему, которые я читал, например, [State of the Art and Prospects for Quantum Computing](#) (2013), впечатляют сочетанием научной строгости и буквально литературного блеска.

«Кончаю, страшно перечесть, стыдом и страхом замираю». Сознаю, конечно, что будь на моём месте другой, список конкретных достижений мог бы заметно отличаться. Только и всего. Существо же дела, как и общие положения о роли теоретиков, были бы, уверен, такими же.

История теоретиков ФТИ была бы неполна, если не упомянуть два громких дела, М. П. Казачкова и Р. Ф. Казаринова, за которые все их коллеги были подвергнуты публичному шельмованию. Они были бы и коллективно наказаны, не вступись сами, энергично и твёрдо, за себя. Казачкова судили в 1975, и приговорили за измену родине к тюремному заключению сроком на 15 лет, которые он провёл в тюрьме и лагере полностью, как говорят, «от звонка до звонка». Поскольку его обвинили в передаче из СССР в США секрета гамма-лазера, который не создан и по сей день, очевидно, что это обвинение было несостоятельно.

Казаринов был лишён научных степеней и званий также в злополучном 1975, поскольку его жена организовала, кстати, против его воли, у них на квартире выставку «незаконных» художников. Детали здесь не важны, и обвинение было бы совсем курьёзно, не имей оно государственной силы, и не трепли оно столько нервов «обвиняемому» и его коллегам. Здесь опять-таки стоит вспомнить о необходимости злопамятности, которая в некоторых случаях полезней и уместней памяти короткой.

*\*\**

Многое изменилось после 1958 в жизни теоретиков в ФТИ. Канул в прошлое «теоретический загон». Исчезли из коридоров второго этажа главного здания группки энергично жестикулирующих и громко спорящих странноватых молодых людей. Вместо них в коридорах чинно движутся в основном управленческие чиновники.

Теоретиков не посылают в колхоз (Фото 1), и они могут спокойно посидеть в рабочее время на лавочке, притом в хорошем парке (Фото 15).

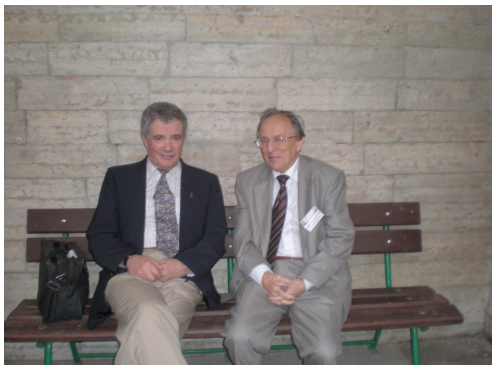

С глаз долой – из сердца вон. Коридор теперь выглядит гораздо лучше, о чём говорит Фото 16. Несколько теоретиков по-прежнему там, но уже в виде молчащих портретов на стенах. За особые научные заслуги не только в виде портрета, а в живую, в Главном здании упомянутых в данном тексте Варшаловича и Яковлева с коллективом разместили на антресолях, но не на кухне, куда классики поселили ничейную бабушку». Остальных же из «загона» переместили в другое здание, в большие комнаты, которые выходят в новые коридоры. Теперь ни нужды, ни охоты гулять в

Фото 15. Варшалович и автор на конференции в г. Пушкин, 2008

них нет. Один из коридоров представлен на Фото 17, и он тоже, как писал поэт, кончается стенкой, однако не в грозном, а обыденном смысле этого слова.

В результате, теоретики теперь не толкутся, где попало, вызывая справедливое раздражение у сотрудников Отдела кадров, и других очень важных отделов.

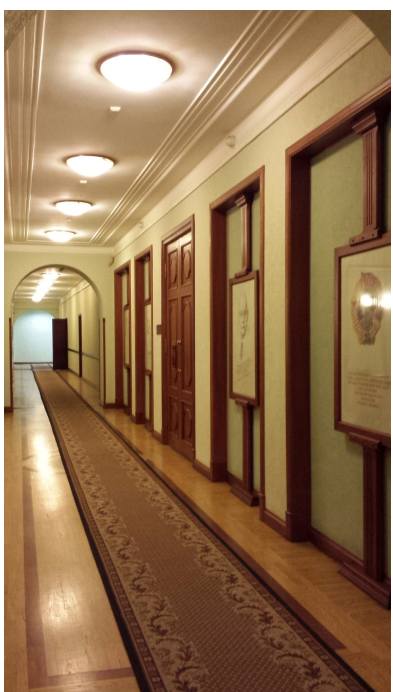

Подразумевается, что процесс творчества протекает за видимыми на Фото 5 дверями, в то время, которое предусмотрено для этого строгими последними решениями высшего руководства бывших учреждений РАН, так называемого ФАНО, с ударением на втором слоге. Прошу не путать с упомянутым в тексте покойным проф. У. Фано! Не знаю, как кому, но мне сегодняшний коридор теоретиков по ширине и уюту

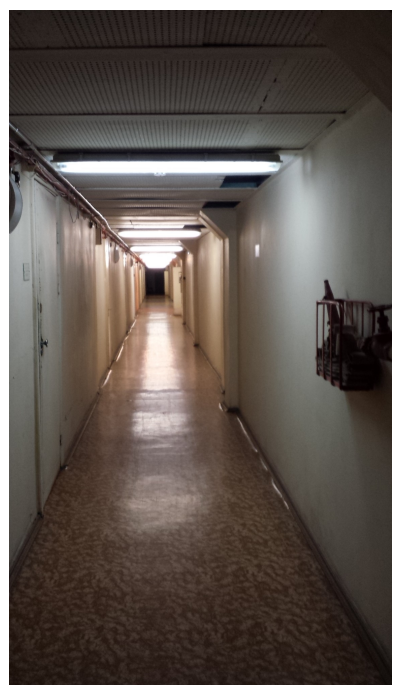

Фото 17. «Теоретический» коридор

Фото 16. Главный коридор с портретами великих.

напоминает тот, что существует в «Крестах», хотя сам я там и не был.

Резко изменились условия оплаты труда – вместо жалких двухсот рублей (в среднем), администрация теперь выплачивает теоретику целых 25 тысяч, опять же в среднем. Так что необходимые условия для работы обеспечены. Впрочем, теоретик от этих условий зависит мало – он заражён желанием работать, устойчивым, и тщательно сохраняемым в организме, как вирус, например, герпеса. А потому дело шло, идёт, и будет идти всегда, что с этим ни делай, простите за невольный каламбур.

Нет, совсем неправ был Мефистофель, говоривший когда-то растлеваемому им студенту: «Теория, мой друг, суха». Приведенные выше результаты явно говорят об обратном.

*Санкт-Петербург*